\documentstyle[aps,epsf]{revtex}  

\def \fplus       {f_{+}}
\def \fs          {f_{S}}
\def \ft          {f_{T}}
\def \lamplus     {\lambda_{+}}
\def \Epi         {E_{\pi}}
\def \Ee          {E_{e}}

\def \Tpi         {T_{\pi}}
\def \Te          {T_{e}}

\def \mpi         {m_{\pi}}
\def \me          {m_{e}}
\begin{document}
\baselineskip 14pt
\title{Scalar and Tensor Couplings in Kaon Decays}
\author{Richard J. Tesarek\footnote{\tt email address:  tesarek@fnal.gov}}
\address{Department of Physics and Astronomy, Rutgers University}
%
\maketitle              

\begin{abstract}
 In the past few years charged kaon experiments have indicated 
 possible scalar and tensor couplings in semileptonic kaon 
 decays($K\rightarrow \pi e \nu$).  These couplings, if correct, 
 are not predicted by the Standard Model and may indicate the onset 
 of new physics.  We present a summary of the existing data and 
 a new, precision  measurement of these couplings in the neutral kaon 
 system based on a 3\% subset of the data taken by the KTeV (E799) 
 experiment at Fermilab.
\end{abstract}

\section{Introduction}
\begin{figure}[ht]
  \centerline{
   \epsfxsize=5.0truein\epsfysize=1.88truein 
   \epsfbox{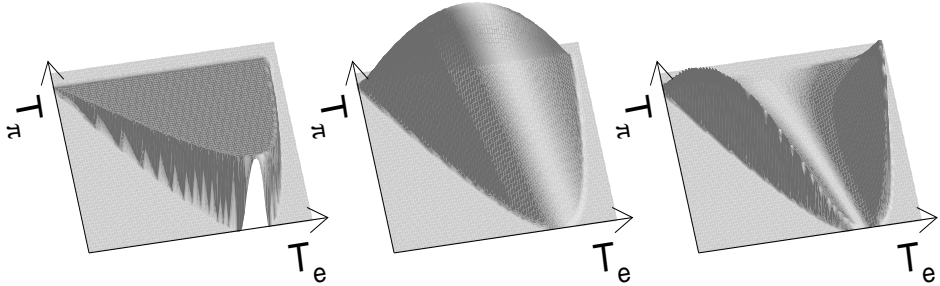}}
   \vspace{0.15in}
  \caption[]{\label{ke3dyn}
  {\small Ke3 Dalitz plots showing the contributions from scalar(left), 
  vector(center) and tensor(right) couplings in the decay.}}
\end{figure}

From relativistic quantum mechanics one learns that the Lorentz 
structure of an interaction is determined by the vertex factors 
or couplings.  The Lorentz structure manifests itself in the 
interaction dynamics or the square of the transition amplitude
and is experimentally accessible.  In semileptonic kaon decays 
involving electrons, $K\rightarrow \pi e \nu$(Ke3), the most 
general form of the decay rate, consistent with the Dirac 
equation and left handed, massless neutrinos is~\cite{ke3_matrix}:
\begin{equation}
 {\cal M} = \left|V_{us}\right| ^2
            \left\{A\cdot |\fs|^2 + B\cdot |\fplus|^2 + C\cdot|\ft|^2 + 
            \;{\rm cross\; terms}\right\} ,
\end{equation}
where $V_{us}$ is the Cabibbo-Kobayashi-Maskawa matrix element~\cite{ckm}
for the $u$ to $s$ quark transition, $A$, $B$ and $C$ are terms 
which depend only on kinematics and $\fs$, $\fplus$ and $\ft$ are 
scalar, vector and tensor form factors, respectively.  The form 
factors parameterize the $K\rightarrow\pi$ hadronic transition and, 
in general, depend on the momentum transferred to the final state 
leptons ($q^2$).  Thus the form factors provide information not 
only on the soft QCD structure of the kaon, but the Lorentz structure 
of the decay.  By analyzing the phase space density or Dalitz plot, 
one may extract these form factors and their $q^2$ dependence which
empirically takes the form~\cite{ke3ff:theory,kpe3exp:full}:
\begin{eqnarray}
f_{i}(q^2) = f_{i}(0)\left( 1 + \lambda_{i}\frac{q^2}{\mpi^2}\right),
  &\quad\quad i=S,+,T
\end{eqnarray}
where $\mpi$ is the pion mass.  The traditional Dalitz plot variables 
are the kinetic energy of the pion ($\Tpi = \Epi - \mpi$) and electron 
($\Te = \Ee - \me$) measured in the kaon rest frame.  
Figure~\ref{ke3dyn} shows how the three different couplings effect 
the shape of the Dalitz plot.
   \vspace{-0.35in}

Since these decays proceed through a virtual $W^{\pm}$ exchange
in the Standard Model, one would expect purely vector couplings.  
However, previous experiments examining the form factors of charged 
Ke3 decays ($K^{\pm} \rightarrow \pi^0 e^{\pm} \nu$) indicate anomalous 
scalar and tensor 
couplings~\cite{kpe3exp:akimenko,kpe3exp:braun,kpe3exp:steiner}.  
The Particle Data 
Group's combined results for the scalar and tensor form factors differ 
from the Standard Model prediction of zero with a significance of 
3.6 and 3.4 standard deviations, respectively.  Previous results from 
neutral Ke3 decays ($K_L^0\rightarrow \pi^{\pm} e^{\mp} \nu$) yield 
only 68\% confidence level upper limits on the scalar and tensor 
couplings~\cite{k0e3exp:blumenthal} and are insufficient to decisively 
confirm or refute the charged Ke3 results. 

In addition to the study of scalar and tensor couplings one may 
extract the $q^2$ dependence of the form factors.  A precise measure
of the $q^2$ dependence of these form factors would allow for a more
precise determination of $V_{us}$.

\section{KTeV}
\begin{figure}[ht]	
  \centerline{
   \epsfxsize=5.50truein\epsfysize=2.50truein 
   \epsfbox{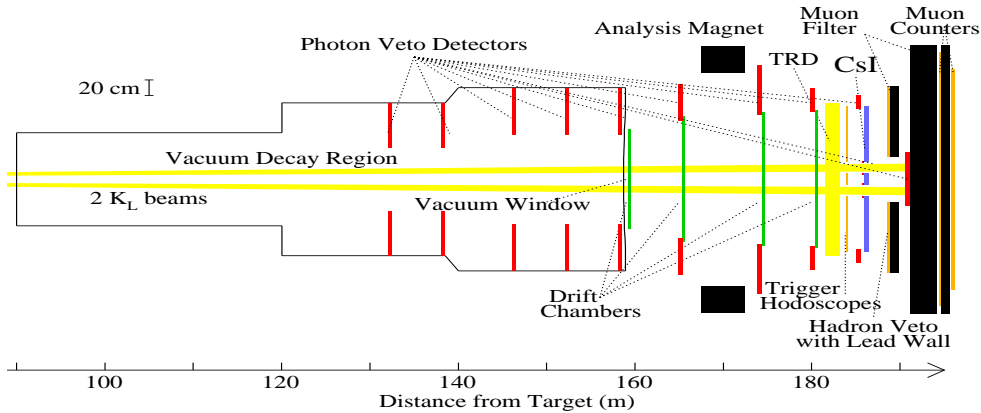}}   
  \vskip -.2 cm
  \caption[]{\label{detector}
   {\small The KTeV Detector as configured for E799.}
  }
\end{figure}

KTeV is a research program at Fermilab comprised of two experiments 
to study rare kaon decay processes (E799) and measure various CP 
violating parameters in the neutral kaon system (E832).   Approximately 
80 collaborators from 12 institutions in the US and Japan~\cite{ktev} 
assembled the KTeV apparatus and took part in data collection and analysis
during the Fermilab 1996-1997 fixed target run.  The KTeV apparatus 
in the E799 configuration is shown in Figure~\ref{detector}.

\section{Reconstruction and Analysis}
The data presented here were taken by the KTeV experiment running 
in the E799 configuration (see figure~\ref{detector}) and consisted
of approximately 7.6 million two track, minimum bias events representing
approximately 3\% of the entire E799 data sample.   The trigger for 
these events required hits in the trigger hodoscopes and drift chambers 
consistent with two charged particle tracks.  

Offline, a Ke3 sample was selected by requiring exactly 2~tracks 
in the fiducial regions of the detector.  The tracks were further
required to form a vertex in the beam region of the vacuum decay volume.
Backgrounds from other decay processes were restricted by requiring
little ($< 100$~MeV) activity in the photon veto detectors and 
no in-time activity in the muon hodoscopes.  Electrons were 
identified from hadrons by the ratio of the measured energy deposited 
in the electromagnetic calorimeter divided by the momentum measured 
in the spectrometer ($E/p$).  The analysis required one electron 
($|E/p -1| < 0.05$) and one hadron ($E/p < 0.80$) assumed to be a 
pion. Additional quality cuts on the tracks and electromagnetic shower 
shape helped to suppress possible backgrounds.  These requirements 
identified a Ke3 candidate sample of approximately 1.4 million events.

\begin{figure}[ht]
  \centerline{
   \epsfxsize=6.0truein\epsfysize=2.75truein 
   \epsfbox{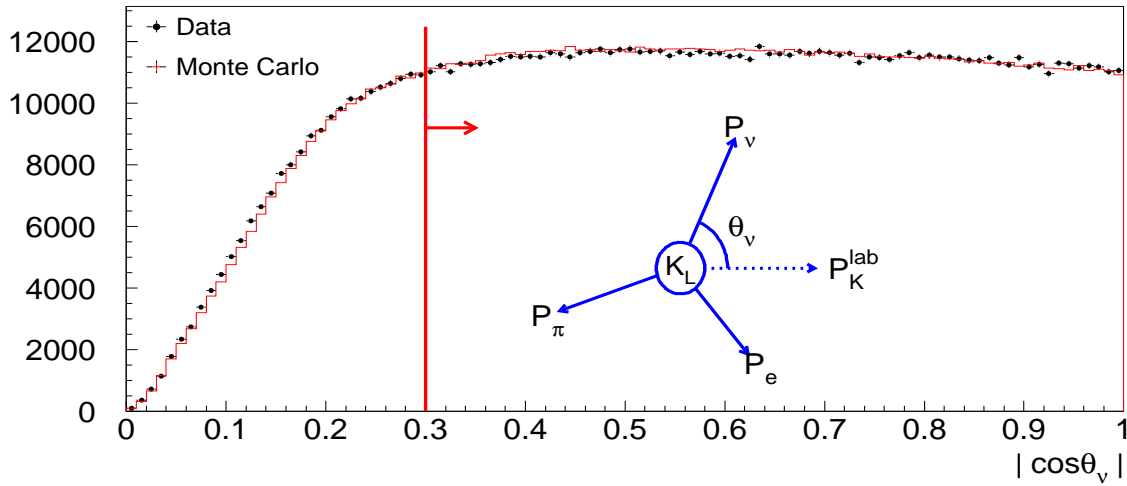}}
  \vskip -.2 cm
  \caption[]{\label{costheta}
   {\small Neutrino angular distribution for Ke3 candidates for data(dots) 
   and a Monte Carlo simulation (line).  Uncertainties are indicated
   by vertical bars for both the data and Monte Carlo.  The inset shows
   a cartoon defining $\theta_{\nu}$.}}
\end{figure}

Because the Ke3 phase space variables are measured in the kaon rest 
frame a good value of the kaon momentum is required to get the 
relativistic boost correct.  Combining the vertex and target locations 
with the measured particle momenta and assuming the decay products 
originate from a kaon allows one to solve for the parent (kaon) 
momentum up to a quadratic ambiguity.  In the kaon rest frame, this 
ambiguity amounts to the lack of knowledge of the sign of the component 
of the neutrino momentum parallel to the kaon flight direction 
($P_{\nu ||}$).  The angle between the neutrino momentum in the kaon
rest frame and the kaon direction may be obtained from the equation:
\begin{equation}
  \cos^2{\theta_{\nu}} = \frac{P_{\nu}^2 - P_T^2}{P_{\nu}^2},
\end{equation}
where $P_{\nu}$ is the magnitude of the missing momentum assuming a
kaon decay and $P_T^2$ is the transverse momentum of the charged 
tracks relative to the line defined by the target and decay vertex.
Figure~\ref{costheta} compares the distribution of $|\cos{\theta_{\nu}}|$ 
from data with a Monte Carlo simulation.  The roll off near 
$\theta_{\nu} = \pi/2$ comes from solutions where $(P_{\nu ||})^2 < 0$ 
and are the result of mismeasurements due to detector resolution, radiation 
and scattering of the parent kaon in collimators upstream of the decay 
volume.  These events were reconstructed by assuming $P_{\nu ||}=0$.  

To calculate the Dalitz plot variables, the electron and pion momentum
were boosted the two candidate kaon rest frames using the high and low
energy solutions to the kaon momentum.  These results were then averaged
to determine a reconstructed electron and pion energy.  By making the 
requirements on the decay kinematics discussed above, the RMS on the 
Dalitz plot variables was reduced from 14~MeV before the cut to 5~MeV.
The final data sample is also reduced to approximately 280,000 events.

The dominant backgrounds to Ke3 decays generally arise from kaon decays
involving charged pions one of which showers in the calorimeter and is 
subsequently missidentified as an electron.  The major background 
sources are from $K_L\rightarrow\pi^{\pm}\mu{\mp}\nu$ 
($30.9\pm 1.5$~events) and  $K_L\rightarrow\pi^+\pi^-\pi^0$ 
($10.0\pm 1.6$~events).  Backgrounds from hyperon decay were also 
considered and contribute $0.54\pm 0.19$~events ($\Lambda\rightarrow p\pi^-$).
Monte Carlo statistics dominate the uncertainties in the background 
estimates.  The total background was estimated at $41.5\pm 2.2$~events 
from all processes and found to be approximately flat across the 
Dalitz plot.

Effects due to electromagnetic radiation have been shown to be
large~\cite{ke3gth:ginsberg,ke3gth:fearing} in some regions of the 
Dalitz plot and must be taken into account.  Radiative effects of 
order $\alpha^2$ 
may be broken down into two classes, a real photon in the final 
state and interference between the tree level diagram and diagrams
involving one virtual photon.  Detailed studies comparing pion-electron
invariant mass($M_{e\pi}$) between data and Monte Carlo show good
agreement when all effects are correctly modeled.  An infrared cutoff 
of 1~KeV was used for the in the simulation of radiative effects.  
This cutoff was varied over six orders of magnitude and no significant 
change of the $M_{e\pi}$ distribution was observed indicating 
cancellation of the different components to high degree.  

The Ke3 data sample was then condensed into a Dalitz plot with 
$5\times 5\;{\rm MeV^2}$ binning.  This Dalitz plot was corrected
bin-by-bin for acceptance and radiative effects involving a real 
photon.  Radiative effects involving a virtual photon were included 
in subsequent fits to the Dalitz plot shape.  The parameters in the 
fit were the linear coefficient of the vector form factor $q^2$ dependence,
$\lamplus$, and the ratio of form factors, $\fs/\fplus$ and $\ft/\fplus$.

\section{Fitting the Data}
In order to avoid personal biases in the analysis, unknown random 
offsets were added to the parameters in the fitting program.  This 
allowed for the detailed study of various detector and reconstruction 
effects without knowledge of the true result.  The analysis strategy 
thus allowed improvements in the detector simulation and reconstruction.
An {\em a priori} requirement that the chisquared confidence level had
to be 10\% or better and detailed comparisons between data and Monte Carlo 
simulations of various distributions had to agree.  In addition systematic
uncertainties were evaluated and reduced until they were of comparable
size or smaller than the statistical uncertainty of the fit parameters.

After all analysis, the fit returned $\chi^2=621$ for 649 degrees 
of freedom (78\% CL).  Table~\ref{syst_summary} summarizes the dominant 
systematic uncertainties for each of the fit parameters and compares
them with the statistical uncertainty returned from the fit.
\begin{table}[tb]
  \caption{\label{syst_summary}{\small
       Summary of dominant systematic uncertainties for the Ke3 form
       factor fit parameters.}}
  \begin{tabular}{lrrr} 
   Effect & $\lamplus$ ($10^{-4}$ units)
          & $\frac{\fs}{\fplus}$ ($10^{-3}$ units) 
          & $\frac{ft}{\fplus}$ ($10^{-2}$ units) \\ 
 \tableline 
   Chamber Effic.               
     & 1.4 
     & 2.8  
     & 3.2 \\
   Acceptance
     & 9.2 
     & 0.70 
     & 2.0 \\
   Resolution \& Radiation
     & $^{+0.3}_{-1.3}$ 
     & 0.85 
     & 0.7 \\
   $P_K$ spectrum               
     &  6.0 
     & 0.75 
     & 0.5 \\
   $P_K$ scale
     &  3.7 
     & $<0.06$ 
     & 1.2 \\
   BR($\rm Ke3\gamma$)          
     &  3.3 
     & $<0.06$ 
     & $^{+0.8}_{-0.4}$ \\
 \tableline  
   Total systematic uncertainty
     & 12.
     &  3.1
     &  4.2 \\
 \tableline
   Statistical uncertainty
     & 8.2
     & $^{+8.5}_{-12}$
     & $^{+5.2}_{-12}$
  \end{tabular}
\end{table}

\begin{figure}[ht]
  \centerline{
   \epsfxsize=6.0truein\epsfysize=3.0truein
   \epsfbox{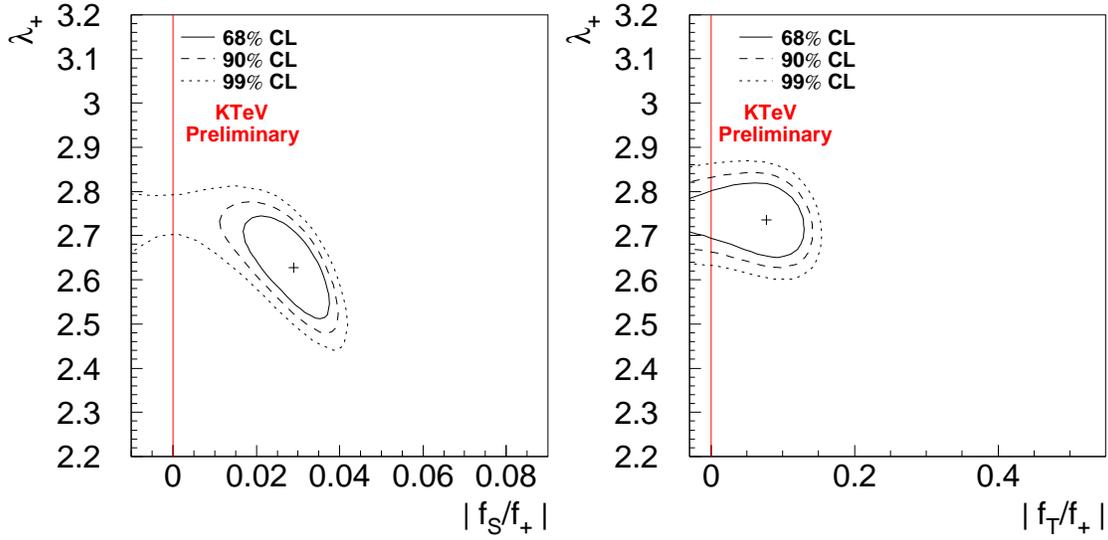}}
  \vskip -.2 cm
  \caption[]{\label{chisqcont}
   {\small  Preliminary chisquared contours in the $\lamplus$ vs 
    $|\fs/\fplus|$ (left)
    and $\lamplus$ vs $|\fs/\fplus|$ (right) planes illustrating the
    correlation between $\lamplus$ and $|\fs/\fplus|$.  A relative
    phase of $180^{\circ}$ is imposed between the vector and scalar or 
    tensor form factor to account for the negative sign obtained in the
    fits.
   }}
\end{figure}

\begin{figure}[ht]
  \vspace{-0.5truein}
  \centerline{
   \epsfxsize=4.75truein\epsfysize=5.2truein 
   \epsfbox{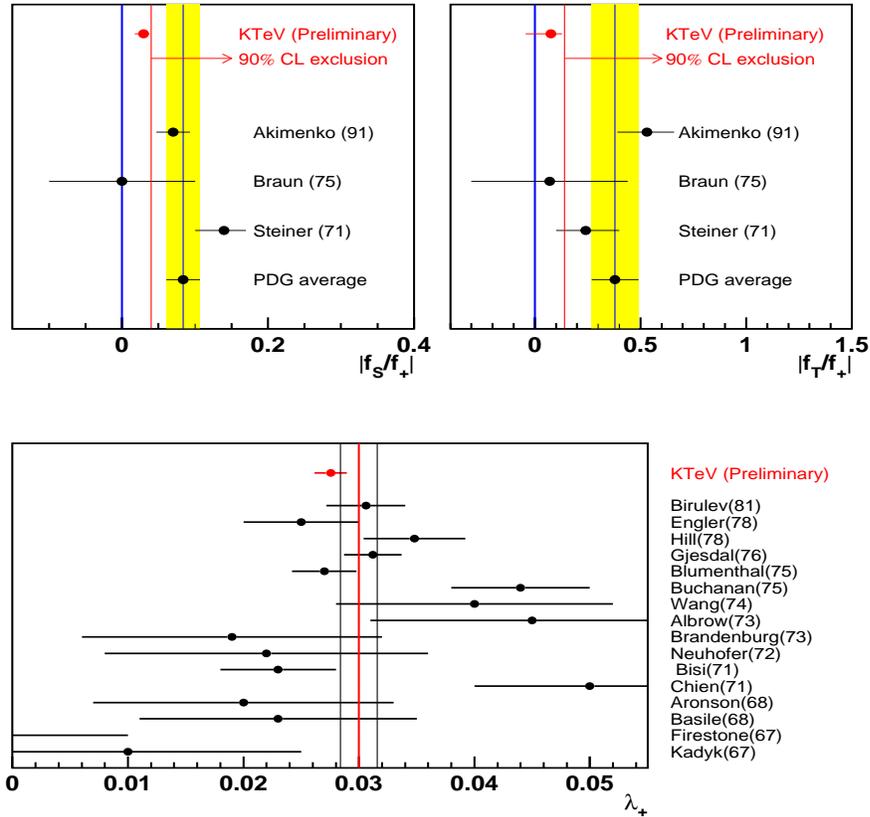}}
  \vskip -.2 cm
  \caption[]{\label{results}
   {\small Preliminary KTeV results for $|\fs/\fplus|$(top left), 
    $|\ft/\fplus|$(top right) form factor ratios and the linear 
    coefficient of the $q^2$ dependence($\lamplus$) for the vector form 
    factor(bottom) compared with previous experiments.
    Vertical bands on each plot represent the Particle Data Group world 
    average.
   }}
\end{figure}

The Dalitz plot was then re-fit with the random offsets removed.  Fits 
were performed assuming a vector, vector+scalar and vector+tensor 
hypothesis separately with the results summarized in Table~\ref{fitresults}.
Figure~\ref{chisqcont} shows chisquared contours for the vector+scalar 
and vector+tensor fit hypotheses.  From these data one may set a 
preliminary 90\% confidence level upper limit on the tensor form 
factor of $|\ft/\fplus|<0.14$.  Because the scalar form factor differs
from zero by little more than $2\sigma$ we quote both a positive value
of $\fs/\fplus = -0.0290^{+0.012}_{-0.0085}(stat)\pm 0.0031(syst)$
and a  90\% CL upper limit of $|\fs/\fplus|<0.04$.  Finally, the 
linear coefficient of the $q^2$ dependence form the vector form factor 
yields the value: $\lamplus = 0.02748\pm 0.00082(stat)\pm 0.0012(syst)$.
These results are summarized and compared with previous experimental 
results~\cite{k0e3exp:full} in Figure~\ref{results}.
\begin{table}[tb]
  \caption{\label{fitresults}{\small
       Summary of preliminary Dalitz plot fit results assuming 
       different hypotheses.}}
  \begin{tabular}{lcccc} 
   Fit Hypothesis & $\lamplus$ 
     & $\frac{\fs}{\fplus}$ 
     & $\frac{ft}{\fplus}$ 
     & $\chi^2/$D.O.F \\ 
 \tableline 
   Vector 
     & $0.02748\pm 0.00082$
     &  --
     &  --
     &  625/651\\
   Vector + Scalar
     & $0.02628\pm 0.0012$
     & $-0.0290^{+0.012}_{-0.0085}$
     &  --
     & 622/650 \\
   Vector + Tensor
     & $0.02735\pm 0.00084$ 
     &  --
     & $-0.0777^{+0.12}_{-0.052}$
     & 624/650 \\
  \end{tabular}
\vspace{-0.25in}
\end{table}
\section{Summary and Conclusions}
Based on a 3\% sample of data taken with the KTeV detector as part
of the E799 experiment, we set preliminary 90\% upper limits on 
possible scalar and tensor couplings in $K\rightarrow\pi^{\pm} e^{\mp}\nu$
decays of $|\fs/\fplus| < 0.04$~(90\% CL) and $|\ft/\fplus| < 0.14$~(90\% CL).
These results represent significant improvements over previously published
results and are inconsistent with the results obtained in $K^+$ decays.
In addition to the coupling results we obtain a preliminary measurement of
the linear coefficient in the vector form factor $q^2$ dependence:
$\lamplus = 0.0275 \pm 0.0008_{\rm stat} \pm 0.0012_{\rm syst}$.



\begin{references}  
 \bibitem{ke3_matrix}
         M.V. Chizhov, eprint {\bf hep-ph/9511287}.
 \bibitem{ckm}
         N. Cabibbo, Phys. Rev. Lett. {\bf 10}, 531 (1963).\\
         M. Kobayashi, T. Maskawa, Prog. Theor. Phys. {\bf 49}, 652 (1973).
 \bibitem{ke3ff:theory}
         L.M. Chounet, J.M. Gaillard, M.K. Gaillard, Phys. Reports,
         {\bf 4C}, 199 (1972).
 \bibitem{kpe3exp:full}
	 C. Caso, {\it et al.,} Eur. Phys. Jour. {\bf C3,} 452 (1998).
 \bibitem{kpe3exp:akimenko}
         Akimenko, {\it et al.,} Phys. Lett. {\bf B259,} 255 (1991). 
 \bibitem{kpe3exp:braun}
         Braun, {\it et al.,} Nucl. Phys. {\bf B89,} 210 (1975). 
 \bibitem{kpe3exp:steiner}
         Steiner,{\it et al.,} Phys. Lett. {\bf B36,} 521 (1971). 
 \bibitem{k0e3exp:blumenthal}
         Blumenthal, {\it et al.,} Phys. Rev. Lett. {\bf 34,} 164 (1975).
 \bibitem{ktev} 
         KTeV Collaboration: 
              Arizona, UCLA, UCSD, Chicago, Colorado, Elmhurst, Fermilab,
              Osaka, Rice, Rutgers, Virginia, Wisconsin
 \bibitem{ke3gth:ginsberg}
         E. S. Ginsberg, Phys. Rev. {\bf 171,} 1675 (1968); 
                         Phys. Rev. {\bf 175,} 2169 (1968).
 \bibitem{ke3gth:fearing}
         H.W. Fearing, E. Fischbach, J. Smith, Phys. Rev. 
                         {\bf D2} 542 (1970).
 \bibitem{k0e3exp:full}
	 C. Caso, {\it et al.,} Eur. Phys. Jour. {\bf C3,} 463 (1998).
 \end{references}
\end{document}